# Vitamin D sensitivity to the immune responses and autoimmunity: A chemical network model study

**Susmita Roy, Krishna Shrinivas and Biman Bagchi***

**SSCU, Indian Institute of Science, Bangalore 560012, India.**

**(Email:** profbiman@gmail.com**)**

## *Abstract*

**Although Vitamin D is believed to be involved in a large number of immune responses our understanding of these processes at cellular level remained at infancy. We develop and solve a coarse grained kinetic network model to quantify the effects of variation of vitamin D on human immunity. The system of equations accounts for known inter-relation between active and inactive vitamin D, antigen presenting cells, effector T cells, regulatory T cells and pathogen. Both time dependent and steady state solutions are obtained. The time dependent solution of the system of equations reveals that the immune response is rather strongly regulated in presence of vitamin D. We found quantitatively that lower than optimum levels of concentration of active vitamin D correspond to weak regulation where, once a pathogen/antigen enters the body, the nature of the immune response would be less regulatory and hence more aggressive, or inflammatory. The steady state solution of our model shows that vitamin D enhances the tolerance level of immune system, thereby increasing resistance to autoimmune diseases. Our model and accompanied numerical analyses reveal another important aspect of immunity: While extremely low levels of vitamin D could lead to increased risk of autoimmune responses, an overdose (toxic) level would give rise to too large a tolerant response, leading to the increased risk of tumors and cancerous cell growth.**

## Introduction

Beyond the well-known function in calcium metabolism it is now clear that vitamin D deficiency can cause several chronic diseases including autoimmune diseases and cancer. Spurred by modern epidemiologic studies, efforts in the last two decades are being directed to understand the origin of non-classical immunomodulatory responses, triggered by active 1, 25-dihydroxy vitamin D. Beyond its established classical function in calcium metabolism, studies on vitamin-D is now progressively focused for its pleiotropic actions. [1-6]

Several experimental and clinical studies have revealed that endogenously produced active vitamin D ($1,25(OH)_2D$) in macrophages enhances the production rate of anti-microbial peptides (cathelicidin, β-defensins, etc), to promote the innate immunity. [7] Subsequently, the conversion of 25-$D_3$ into functional 1, 25-$D_3$ [D*] exerts potent effect on adaptive immune system. It apparently slows down the functions of antigen-presenting cells (APCs), particularly of Dendritic cells (DC). [8] While active form of vitamin D severely affects T cell activation, proliferation and differentiation, it facilitates the production of regulatory T (Treg) cells that can indeed function as an effective immune controller. [9, 10]

Vitamin D mediated infections remedial therapies have been followed over past 150 years. Since early 1900s, usage of cod-liver oil and UV light became widely recognized as the essential sources of vitamin-D. Therapeutic use of vitamin D first drew attention in 1849, when William found cord- liver oil in curing over 400 tuberculosis (TB) patients. [11] After a long 50 years gap, Niels Finsen owned the Nobel prize by highlighting the medicinal approach of UV exposure by which he treated over 800 patients affected by lupus vulgaris (a cutaneous form of TB). [12, 13] In Indian traditional Auyrvedic treatments, use of sunlight to treat and reduce diseases goes back to several thousand years where it is referred to as "Suryavigyan" (Meaning: science of Sunlight).

We now briefly review the biochemistry of the processes involved in vitamin D. Vitamin D, a type of secosteroid can be ingested by humans as cholecalciferol (D3) or ergocalciferol (D2). Humans can also generate Vitamin D by converting cholesterol when exposed to sunlight. In fact, this mechanism not only contributes to a significant percentage of the production of Vitamin D but also in maintaining serum levels through a feedback loop. The hormonally active form of Vitamin D is calcitriol (1, 25-dihydroxyvitamin D3). This generation of the active form is through multiple steps. The 1$^{st}$ step is the hydroxylation of the inactive Vitamin D3 in the liver to 25-hydroxyvitamin D3 (calcidiol). This is then converted to 1, 25-dihydroxyvitamin D3 via the enzyme (cytochrome P450 monooxygenase 25(OH)D 1-α hydroxylase (CYP27B1; 1α(OH)ase) mediated catalysis of 25-hydroxyvitamin D3 . [14] This enzyme is primarily found in the proximal renal tube. But the enzyme itself is upregulated and stabilized in various other cells, including leukocytes, macrophages and dendritic cells in presence of appropriate stimuli. Active vitamin D (D*) also auto-regulates its synthesis by upregulating the enzyme 25-Hydroxyvitamin D3 24-hydroxylase (CYP24A1), which hydoxlyates both Calcitriol and Calcidiol into hormonally inactive forms. This enzyme (CYP24A1) is also a P450 cytochrome. [15]

Most typical immune responses are divided into two arms: Innate response and adaptive response. Innate response involves a generic response against pathogens primarily referred by phagocytes and natural killer cells. In this response, macrophages and monocytes provide the 1$^{st}$ line of defense against the invading pathogens by phagocytosing the pathogen with the help of some receptors (PAMPs). [8] Dendritic cells in the Lymph node (mostly Myeloid Dendritic Cells) serve to break down the pathogenic epitopic peptides and play a key role in activation of the adaptive response. Natural killer cells also kill pathogens, but not by direct phagocytosis of pathogen, but rather by killing virally infected cells. In contrary adaptive response is a specific pathogen based response which is triggered by activation of antigen specific T and B cells in jawed vertebrates. This response, though late to set in, not only serves to mitigate pathogenic growth but also renders the body with immunological memory, which helps in easier control of a particular pathogenic outbreak, if it were to happen again. The processes of somatic hypermutation and some specific (VDJ) recombination help in expressing a vast number of antigen receptors from a small number of genes. This essentially explains how the T and B cells can identify virtually almost any type of antigen. [16]

**Vitamin D and innate response:**

Vitamin D plays a crucial role in the functioning of the innate immune system, both by aiding the killing of pathogens and regulating the process which triggers the adaptive response. An in-depth analysis follows:

Vitamin D is found to be a stimulant in the maturation of monocytes to macrophage-like phagocytosing cells. The antimicrobial activity exhibited by these Macrophages is highly dependent on the presence of active vitamin D (D*). The following is a point wise, sequential explanation of a macrophage response: (i) The pathogens exhibit certain epitopes which are used to recognize a foreign organism that is present. These generic epitopes are called Pathogen Associated Molecular Patterns (PAMPs), which are used by the innate immune system to respond against invading pathogens. (ii) There are Toll-like Receptors (TLR) that are present on the surface of the macrophage cells which help in recognizing these PAMPs. Once this recognition process occurs, a biochemical cascade of events is triggered which ultimately leads to the death of the pathogen. (iii) The immediate response post TLR activation is the NF-κβ activation through the MyD88 coupling pathway. [17] This leads to three major results: Upregulation of the enzyme CYP27B1 in the P450 cytochrome and upregulation of VDR expression. Monokine synthetic response primarily includes release of interleukins 1β and 15. [18] (iv) Once the enzyme CYP27B1 is upregulated, there is increased local production of D* from inactive D3. Since VDR expression is also up-regulated, this also leads to the formation of more D*-V DR complexes. [D*] is regulated in two ways:

(a) Autoregulation: As more D* is produced, the enzyme CYP24A1 is upregulated. This enzyme in turn inhibits formation of D* from inactive D3.

(b) Ligand-Binding: A truncated version of the CYP24A1 enzyme version is also present,i.e. CYP24A1-SV(Splice Variant). Despite being metabolically inactive, it can act as a decoy for D* or inactive D3 by binding with them in the steroid binding pockets. Molecular modeling suggestsscus this Splice Variant preferentially binds to inactive D3 suggesting that it acts as a decoy for inactive D3, reducing its availability to other enzymes. [18, 19]

**Vitamin-D and adaptive response:**

Myeloidic Dendritic Cells (DC) play an important role in activating the adaptive response. They are also known as professional Antigen Presenting Cells (APC). These dendritic cells phagocytose pathogens and present certain epitopes in the Major Histocompatibility Complex (MHC) molecule type II on the surface of the DC. This process through which a dendritic cell phagocytoses a pathogen and presents it on the surface as a MHC class II molecule is referred to as the maturation of the dendritic cell. Naive T cells which are produced in the Thymus gland are activated by coming in contact with a dendritic cell which presents a MHC class II type peptide. This process occurs in the lymph nodes. The process is quite complex, but can be understood as the basis for the activation of the pathogen-specific adaptive response. The basic activation step can be understood by understanding the process of T-cell activation. The chronological analysis of the event is described in the supplementary meterial [20].

D* concentration affects the formation of different type of helper T cells. The most prominent effect of [D*] is in the stimulation of production of IL-10. This leads to dual effect of increasing the Treg cell count but also inhibiting Th1 and Th17 cell proliferation. However the effect of vitamin D on Th2 is not yet clear. The Treg cells are positively regulated through the up-regulation of the CLTA-4 and FOX-P3 markers. [D*] also suppresses production of IFNγ, IL-17, IL-21 and IL-22. This also suppresses proliferation of Th1 and Th17 type effector cells. Furthermore, production of vitamin D is directly enhanced by the formation of the activated effector T cells. This is due to the fact that the CD40 bonding upregulates 1-α expression (CYP27B1) in the dendritic cells. Also, as mentioned above, vitamin D can down-regulate itself by up-regulating the enzyme CYP24A1 in the P450 cytochrome. [21-24]

*DCs are the central linker between innate and adaptive immune systems*. While DC activation occurs through pattern-recognition receptor referred as innate response, in turn trigger the activation of adaptive immune network in a sequentially manner. [24]

**Vitamin D and autoimmune diseases**

Targeted destruction of self-tissues by the overly active immune system causes various types of autoimmune diseases. Autoimmunity is primarily driven by enhanced number of T helper cells (Th1) that attack various self-tissues in the body. It is clear that both genetic and environmental factors affect disease prevalence. However, from past epidemiologic data highlights the link between vitamin D insufficiency and a range of immune-mediated disorders have emerged in parallel with experimental studies on the immunomodulatory properties of vitamin D. In particular the inhibitory effect of vitamin D on the effector T-cell responses and promoting Tregs, may, at least in part, have begun to explain some of these associations. [8, 25-27].

In the present article we develop a theoretical coarse grained kinetic network model model to explore quantitatively the dependence of immunity on vitamin D and how it could play an important role in reducing the risk of auto-immune diseases and cancer. We analyze several immunemodulatory issues that essentially controlled by vitamin D including both innate and adaptive responses as articulated in several experimental reports and reviews. Though there are numerous other complex biochemical reactions involved in the immonological pathways, we have considered only a certain number of them that involve important interaction with vitamin-D. We also address the concern for optimal vitamin D intake raised by World Health Organization‘s international agency for reasearch on cancer. Present study especially suggest that inhibitory action exerted by regulatory T-cell induced by vitamin D and by vitamin D itself on adaptive response play an important role in prevention of autoimmune diseases.

The present approach of network kinetic model building bears strong resemblance to similar methods adopted routinely in the study of kinetics proof reading [28, 29] and in also enzyme kinetics [30-32]. In all these studies, precise quantitative prediction is hindered by imperfect knowledge about the system parameters; especially values of rate constants are often not available. This lacuna is indeed a source of serious problem not only in study of kinetic proof reading, enzyme kinetics but also, as we find here, in theoretical investigations of immunology. Finally, master equations involved all such problems are solved essentially by employing the method of mean first passage time [29, 33], Gillespie algorithm or straight forward numerical integration. We adopt the last procedure here.

In the present study we have employed the parameter space scanning method to obtain their order of magnitude values. We believe that we are not missing any dynamical character of the system. However the present method we have employed is deterministic as we have ignored the necessary fluctuation evolved in the system.

# Method

## Coarse-graining of the reaction network

To understand the interplay among different types of immune cells, pathogens and the modulatory role of vitamin D, a simple coarse-grained approach needs to be developed. This is because biochemical machineries in the human body are highly coupled with each other and hence leads to enormous complexity. Hence understanding the relation between different machineries involving different types of cell requires high level of detailing at the molecular level. A simpler, albeit cruder version is proposed that accounts for different level of complexities at the molecular levels by coarse-graining them at the cellular level. Subsequently we perform network analysis based on T-cell activation, deactivation ad regulation. Once the developed network appears to be simple enough, a system of coupled differential equations is used to model the system. Even our simpler model involves nine coupled equations which need to be solved numerically and, as shown below, exhibit rich dynamics.

We initially represent a complex, rather detailed model system based on the interaction network. This representation includes DC (dendritic cells) linker that essentially connects the innate and adaptive immune network.  The main proposed model is not at the molecular level encompassing MHC−II, but rather at the cellular level. This model primarily seeks to understand adaptive response activation and effect of vitamin D on tolerance/regulatory nature of the response. Hence we have emphasized over the regulatory function of vitamin D on adaptive immune system considering a constant rate of pathogen killing by innate mechanism with the production of antimicrobial peptides leading the 1st defense against infectious diseases.

The important constituents considered here are the following: (i) pathogen/antigen/epitope/self-Antigen, (ii) naive T-cell, (iii) myeloid dendritic dell in the form of professional antigen presenting cells (APC), (iv) effector/regulatory T cells, (v) vitamin D (25(OH)$D_3$), (vi) vitamin D receptor (VDR) and (vii) the enzyme 25(OH)D3-1a-hydroxylase (CYP27B1) that simultaneously function over inactive vitamin D to form activate vitamin D (1,25$(OH)_2$ D)-VDR protein complex. It is important to emphasize here that we have included the essential features of the adaptive responses mostly based on the interaction network proposed by Powri and Maloy. [34] However they did not consider or include the important role of vitamin-D. In **Figure 1** we have presented the complex interaction network model that includes various components and their inter-relation and regulation involved in the immune system.

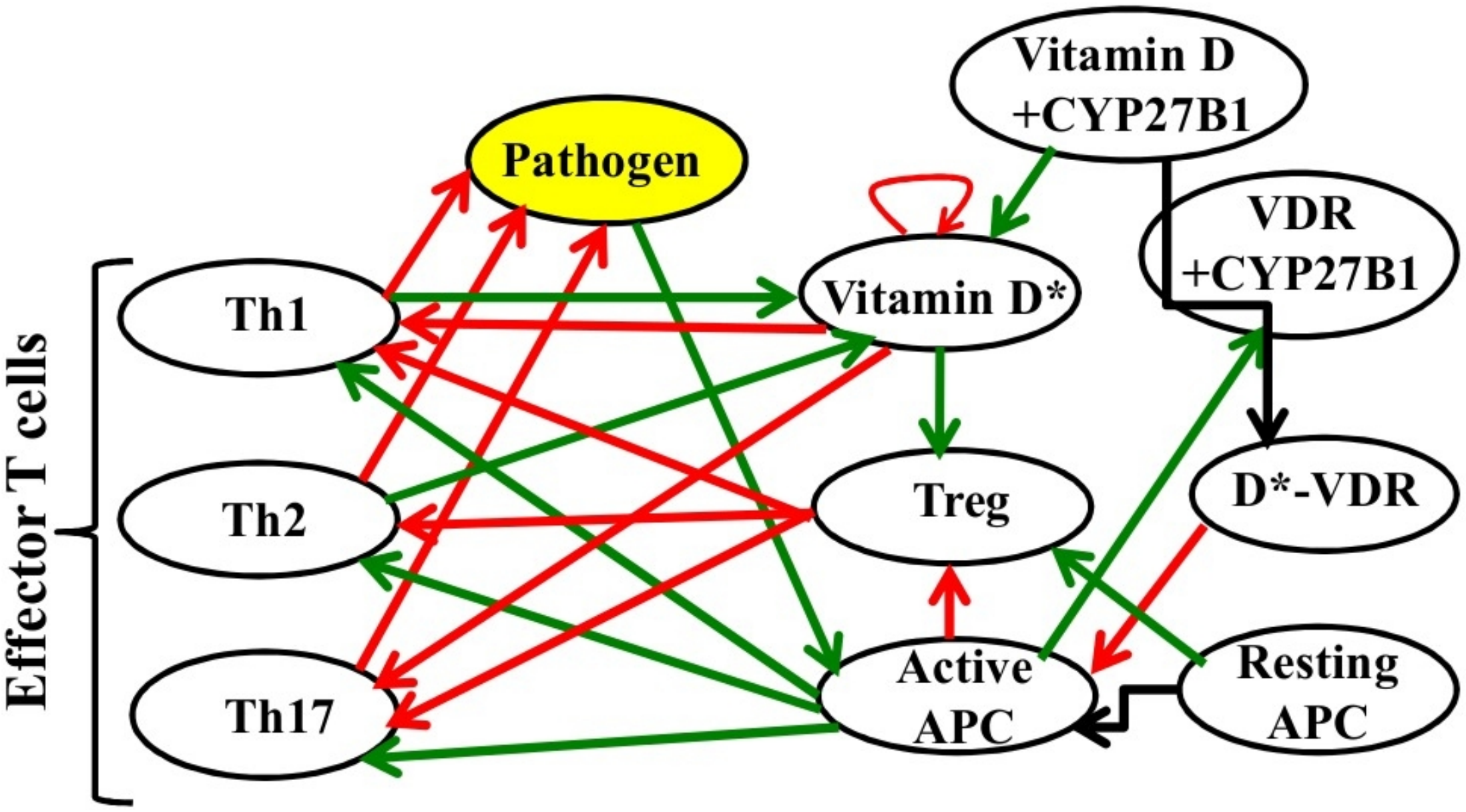

**Figure 1: A complex representation of adaptive response.**

In this network the events are the following: (i) In presence of pathogen APC become stimulated and induce the activation of effector T-cells that ultimately destroy the pathogen. (ii) The enhanced rate of active T-cells production is regulated by both regulatory T-cells (Treg) and active vitamin-D. (iii) In behind regulatory T-cell function is upregulated by active vitamin-D either through membrane initiated interaction or through VDR mediated mechanism. (iv)In addition more regulatory T-cell leads to more resting APC that again leads to more regulatory T-cells. Similarly more effector T-cell leads to more active APC that again leads to more effector T-cells. The self feedback loop is of vitamin D is noted. We have not shown the precursor cells for APC or T-cells in this diagram. *Note that here green arrows stand for upregulation and red arrows for inhibition.*

It is well established that the primary molecular action of 1, $25(OH)_2$ D is to initiate gene transcription by binding to VDR which is a member of the steroid hormone receptor superfamily of ligand-activated transcription factors. VDR therefore is an important factor in 1, $25(OH)_2$ D mediated functions. More detailed information about the molecular activation of VDR and its role in gene transcription can be obtained from the recent review appraised by Pike et al. [35]

In contrary, we found that there are growing evidences that 1, $25(OH)_2$ D also has rapid actions that are not essentially mediated through transcriptional events involving VDR. They are in fact membrane initiated actions. [36]

Our model is essentially based on the following observations:

(i) Myeloid Dendritic Cells (here we call them as antigen presenting cell (APC)) which circulate around the body are the key players involved in triggering the onset of an adaptive response. While pathogen phagcytosed Dendritic cells serve to activate naive T cells into effector T cells, immature Dendritic cells (APC) convert naive T cells into regulatory T cells.

(ii) Naive T cells are circulated and produced in large quantities in the blood stream and using absolute amount of T cells is both ineffective and computationally expensive. Instead, only the antigen-specific naive T cells are considered in the model. Hereafter, when references are made to native T-cells, they only refer to antigen/epitope specific naive T cells.

(iii) Pathogen/Antigen has a birth rate which includes influx and proliferation rates and a death rate similar to decay which incorporates natural cell death and also due to innate response. Besides this it can also be destroyed directly or indirectly through the effect of T-effector cells.

(iv) Active Vitamin D, [D*], has a tight control over the homeostatic production rate that auto-regulates its production by directly upregulating the activity of the P450 cytochrome CY24A1. Also, the T-effector cells release cytokines which upregulate the activity of P450 cytochrome CYP27B1, which in turn increases production of [D*].

(v) While T-effector cells are directly inhibited by the increased production of [D*], Treg cells become up-regulated. In addition, Treg cells inhibit T-effector cells proliferation.

In the next step we work on coarse-graining the interaction network. This is accomplished through making a few simplifying observations and vital assumptions. They are as follows:

(a) Th1, Th2 and Th17 cells are grouped together as Teffector cells.

(b) The circulating, inactive form of vitamin D, 25(OH)D3, is generally used as an indication of vitamin D status. However, in dendritic cells (DC) use of this precursor depends on its uptake by cells and subsequent conversion by the enzyme CYP27B1 into active 1,25(OH)2D3. As 1,25$(OH)_2D_3$, the active form of vitamin D, exerts potent effects on the cells of the immune system and regulating T cell activation, differentiation and migration, we have taken into consideration only the active vitamin D (1,25$(OH)_2D_3$) status in DCs. It is represented as [D*].

(c) $[D*]$ is assumed to have an overall down regulatory effect on Teffector cells.

(d) The T-effector cells are all downregulated by regulatory T cells and all of them kill pathogens. It is assumed that T-effector cells have an overall upregulatory effect on [D*] production. Also noted that, regulatory T cells are promoted by vitamin D and are inhibited as $[MHC - II]$ increases (as Treg are upregulated by immature DC).

**Figure 2** summarizes the whole course graining method.

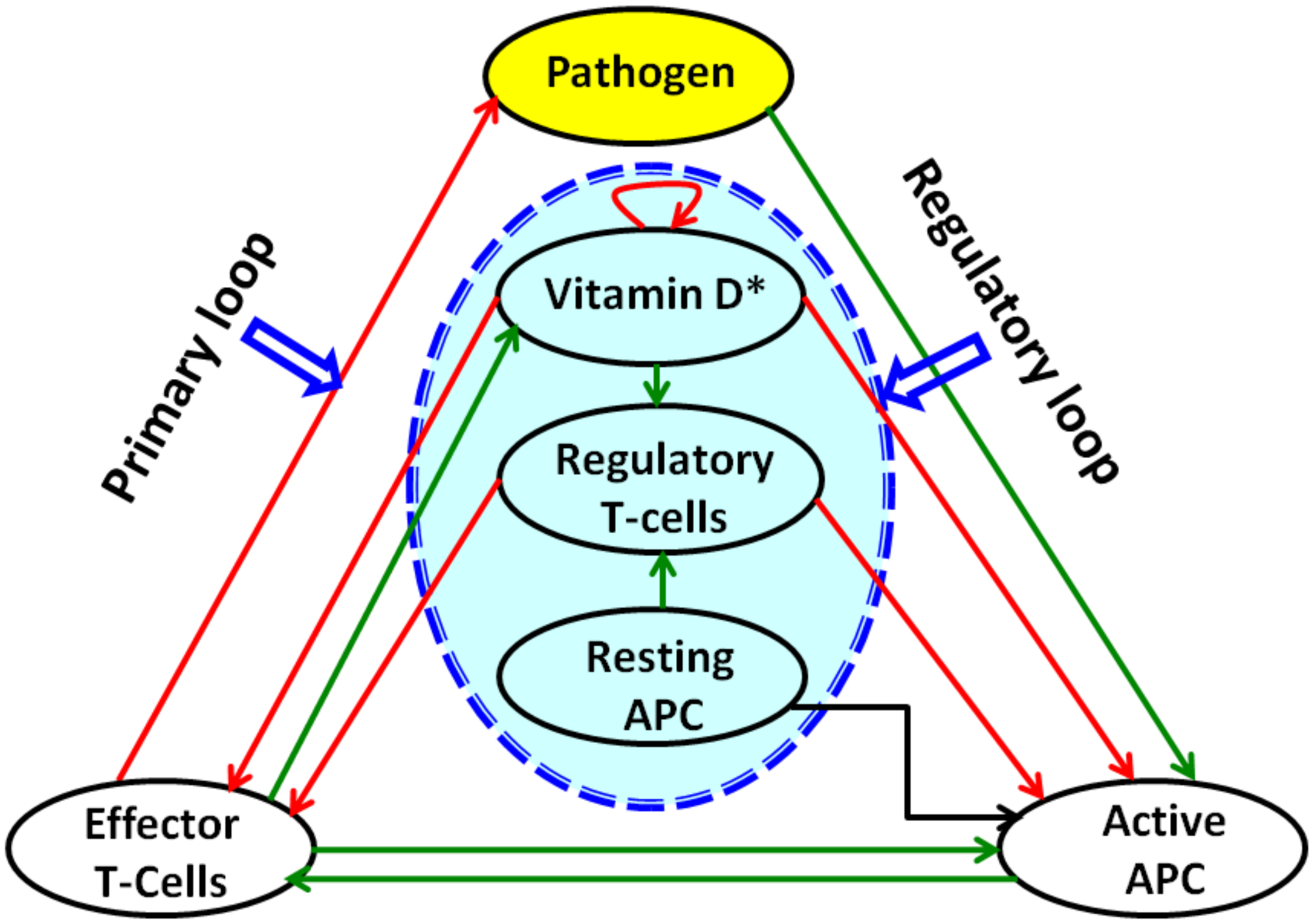

**Figure 2: Coarse-grained network of Figure 1.**

In this diagram for simplicity we consider two loops: (i) Primary loop, (ii) Regulatory loop. Primary loop includes pathogen, active APC and effector T-cells. These components are involved in pathogen killing process. Regulatory loop have resting APC, Treg cells and vitamin-D. While regulatory loop inhibits the overexcited function of primary loop, primary loop promotes the function of regulatory loop through vitamin-D-effector T-cells interaction. Here also green arrows stand for upregulation and red arrows for inhibition and black arrow represents conversion processes.

Now, some important further assumptions before we set about writing the master equations:

(i) All death rates of T-cells are linear in concentrations.

(ii) The transition probabilities are all assumed to be constant with time but may vary from system to system (i.e. here person to person) according to the condition applied.

## Master equations quantifying the reaction network dynamics

Once the following considerations are taken into account, following the coarse-grained network for the system as described above, a set of the master equations are written down for the system.

$$\frac{dP}{dt} = \sigma_P - \pi_P - k_P T_E P$$

$$\frac{dA_{in}}{dt} = \sigma_{in} - k_{inP} A_{in} T_E - m_a A_{in}$$

$$\frac{dA_{res}}{dt} = k_{inP} A_{in} P + k_{aR} T_R A_a + k_{aD*} A_a D^* - k_{aE} A_{res} T_E - k_{inP} A_{res} P - m_a A_a$$

$$\frac{dA_a}{dt} = k_{aE} A_{res} T_E + k_{inP} A_{res} P - k_{aR} T_R A_a - k_{aD*} A_a D^* - m_a A_a$$

$$\frac{dT_N}{dt} = \sigma_{T_N} - k_{aN} A_a T_N - k_{resN} A_{res} T_N - m_N T_N$$

$$\frac{dT_E}{dt} = k_{aN} A_a T_N - k_{ER} T_E T_R - k_{ED*} T_E D* - m_E T_E$$

$$\frac{dT_R}{dt} = k_{resN} A_{res} T_N + k_{RD*} T_R D* - m_R T_R$$

$$\frac{dD}{dt} = \sigma_D - k_{ED} T_E D - m_D D$$

$$\frac{dD*}{dt} = k_{ED} T_E D - m_{D*} D*$$

Where the terms signify as follows:

$k_{ij}$ → Transition probability rates,

$\sigma_k$ → Production rate by body of component $k$ ,

$m_i$ →Overall death rate of component $i$ ,

$P$ → Pathogen/Antigen/Self-Antigen etc,

$A_a$ → Mature Antigen Presenting Cells,

$A_{in}$ → Immature Antigen Presenting Cells,

$T_N$ →naive antigen-specific T cells,

$T_E$ → effector T cells,

$T_R$ →regulatory T cells,

$D$ → Inactive form of Vitamin D3 (**1, 25(OH)D)** in the body

$D*$ →active form of Vitamin D3 (**1, 25(OH)$_2$D)** in the body

## System parameters

The set of nine coupled differential equations is very difficult to solve analytically. Hence numerical methods have to be applied to understand the dynamics of the system. Since accurate rate constants/transition probabilities in most of the cases are unknown here, order of magnitude estimates for many of the values were employed. [37] In addition, the concentration of precursor elements was normalized, so as to reflect manifold change in the production level. Taking typical values as mentioned below (see **Table 1**), the time evaluation of the system and other analyses are performed in the present work.

**Table 1: Basic parameter values (*time duration is taken as "days").**

| Parameter | Symbol | Value |
|---|---|---|
| Reproduction rate of pathogen | $\sigma_P$ | **1** |
| Death rate of pathogen | $\pi_P$ | **1** |
| Birth rate of APC | $\sigma_{in}$ | **0.2** |
| Death rate of APC | $m_a$ | **0.2** |
| Rate of pathogen killing by efffector-T cells | $k_P$ | **10** |
| Rate of APC activation by pathogen | $k_{inP}$ | **variable** |
| Rate of APC reactivation by effector T cells | $k_{aE}$ | **variable** |
| Rate of APC inhibition by regulatory T cells | $k_{aR}$ | **1** |
| Rate of APC inhibition by active vitamin D | $k_{aD*}$ | **$10^{-2}$** |
| Birth rate of Native T cells | $\sigma_{T_N}$ | **1** |
| Rate of differentiation of native T cell to effector T cell induced by | $k_{aN}$ | **1** |
| active APC | | |
| Rate of differentiation of native T cell to regulatory T cell induced by resting APC | $k_{resN}$ | **1** |
| Mortality rate of native T cell | $m_N$ | **0.01** |
| Rate of inhibition of effector T cell by active vitamin D | $k_{ED^*}$ | **1** |
| Rate of inhibition of effector T cell by regulatory T cell | $k_{ER}$ | **0.01** |
| Rate of decay of effector T cells | $m_E$ | **0.1** |
| Rate of regulatory T cell reactivation by active vitamin D | $k_{RD^*}$ | **0.01** |
| Rate of decay of regulatory T cells | $m_R$ | **0.1** |
| Production rate of inactive vitamin D | $\sigma_D$ | **1** |
| Death rate of inactive vitamin D | $m_D$ | **0.1** |
| Rate of reactivation of active vitamin D induced by effector T cells | $k_{ED}$ | **1.0** |
| Rate of deactivation of active vitamin D | $m_{D^*}$ | **1** |

## Results

Under any sort of pathogenic attack, a healthy immune system always aims at killing the foreign antigen by enhancing the proliferation and differentiation rate of its effector T-cells. However, the over exceeding number of effector T-cells often fail to distinguish between body's self and foreign peptides and may start destroying self tissues. Thus a healthy immune system has to operate within a balanced regulation and the concentration of effector T-cells must be controlled so that it can effectively overcome both the pathogenic stimulation and also can resist the exploding growth of effector T-cells. Here comes the role of vitamin-D whose optimum level effectively maintains a balanced immune regulation. Vitamin D motivates regulatory T-cells activity, as well as vitamin D itself is assigned to reduce the hyper activity of APCs and effector T-cells.

In several other model studies only regulatory T-cells are assumed to maintain such regulation. Although there are several experimental and clinical observations revealing the important role of vitamin D and its concentration dependent effect in such regulation, we are not aware of a single model study that has been employed to study the role of vitamin D.

## Effect of Vitamin-D on T-cells regulation: From weak to strong

When human body is exposed to certain pathogen of any weak to moderate strength, our immune system immediately switches on APC activation and effector T-cells activation. The upregulation processes of the regulatory T-cells by resting APC and resting APCs by regulatory T-cells as well as upregulation of effector T-cells by active APC and more APC activation by effector T-cells play a crucial role is in any strongly or weakly regulated state. [34] However the competing behavior of regulatory T-cells and effector T-cells never come to an end in presence of antigen because resting APCs are constantly produced by capturing the antigen and APCs are activated by antigen stimulation. Thus not only the antagonistic interaction between regulatory and effector T-cells activity but their regulation by vitamin-D eventually determine the fate of a disease.

In the present study we have categorized the regulation into three groups: (i) Strong regulaion, (ii) weak regulation and (iii) moderate regulation. To investigate several vitamin-D associated factors we have performed time evolution analysis of each participating element after the pathogen attack to study their long time behavior. We have studied all these three regulation limits both in absence and presence of vitamin-D at different antigenic stimulation. The results are shown in **Figure 3** are a series of curve for all the three regulation limits, both in the presence and absence of vitamin D. The results are quite interesting and we discuss them in more detail below.

In an early study Fouchet and coworker [37] analyzed the steady state values of T-cells in these three regulation regimes and showed an interesting and important bistable region at moderate regulation regime. We have also observed such bistable region in absence of vitamin-D as shown in **Figure 3(e)** where both strong and weak regulation can coexist.

Here we find from **Figure 3(b)** that when system falls under a strong regulation the presence of vitamin D shows negligible change compared to a situation when there is no vitamin D (see **Figure 3(a)**). In contrary we observe that in the presence of even a minute amount of vitamin-D can convert a weak regulation to a strongly regulated state. (See **Figure 3(c)** and **Figure 3(d)**). At moderate regulation regime, vitamin D is found to affect the bistable region (see **Figure 3(e)** and **Figure 3(f)**). In this regime, strong regulation is found to become much stronger, even weak regulation is also found to revert towards a strongly regulated state.

When antigenic stimulation is very high the system always is found to move forward to a weakly regulated state where the effector T-cells are abundant. In any situation of **Figure 3**, we observe the crossover of T-cell population towards the high antigenic stimulation $(k_{inP})$. Depending on the intensity of antigenic stimulation, system mounts on the required effective strong regulation to control the inflammation. This result suggests the important role of vitamin need in switching on such effective regulation.

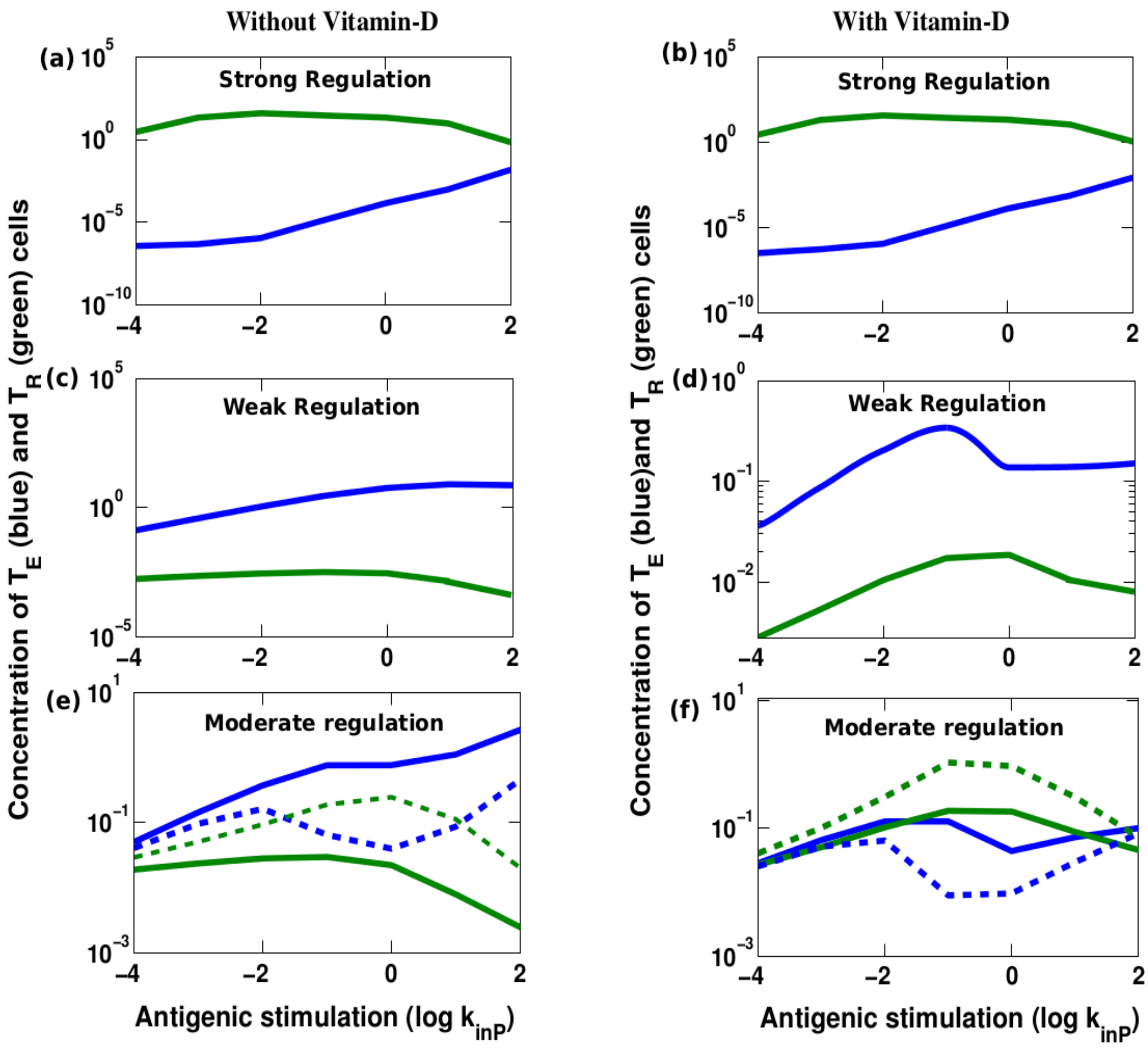

**Figure 3: Variation in T-cell regulation under weak to strong regulation.**

Steady state concentrations of effector T-cells (shown in blue) and regulatory T-cells (shown in green) are plotted against various range of pathogenic stimulation $(k_{inP})$ at the three different APC mediated effector T-cell regulation $(k_{aE})$. We find a stable strongly regulated state $(k_{aE} = 10)$ both (a) in absence of vitamin-D and (b) in presence of vitamin-D. The strong regulation remains strong also in presence of vitamin-D. An example of stable weakly regulated state $(k_{aE} = 10^4)$ both (c) in absence of vitamin-D and (d) in presence of vitamin-D where weakly regulated state turns to be strongly regulated state up to certain pathogenic stimulation $(k_{inP} \approx 10^0 = 1)$. An example of bi-stable state: $(k_{aE} = 10^3)$ (e) where both weakly regulated state (shown in solid line) and strongly regulated state (shown in dashed line) can coexist in absence of vitamin D. (f) In presence of Viatmin D both the stable states correspond to the strongly regulated states. However in the bi-stable situation there always exists an unstable state. Note that strong regulation exerted by vitamin D prolongs for a very high antigenic stimulation. Here we consider the other rate values that are given in Table 1.

## Time evolution of immunological components

### In absence of (or at very low concentration) of Vitamin-D

We observed some interesting results from study of the time evolution analysis of the immunological components in the above mentioned three regions. Here we have presented the dynamical change of elements against time (days) which can quantitatively explain some attributes of the immune responses both in absence (see **Figure 4**) and presence of vitamin D (see **Figure 5**). In **Figure 4(a)** at around day 2, we see that there is a sudden increase in amount of resting APCs (see **Figure 4(d)**), which reaches a peak very quickly. This, as said before, is typically referred to as the onset of the adaptive response. This is in common agreement with most experimental results which suggests that the adaptive response sets in around 2 days (48hours) or so after the pathogenic incursion. [38]

Around the same time resting APCs start getting activated and we see from the graphs in **Figure 4** that the pathogens start dying out at a much faster rate. We now have a huge population of T-effector cells which have been activated from the naive T-cells after APC activation.

We also find that once the resting APC has reached its peak level (around 2$^{nd}$ day), regulatory T-cells start to raise their production. Once this happens, T-effector cells and APCs that have reached a peak are now started to be regulated (around 2$^{nd}$-5$^{th}$ days). But however when pathogen almost die off caused by the tight control exerted by effector T-cells, then resting APCs fall in number and regulatory T-cells also follow the same trend as well. Once the Treg cells fall down significantly body is then no more tolerant of the pathogen.

Also, once steady state has been reached after 100 days or so, we can notice that the steady state values of effector T cells (see **Figure S1**). It has been clearly shown in the supplementary material. This is in accordance with the fact that these cells are antigen specific cells and once the pathogen is suppressed, the body creates an immunological memory of the pathogen, which corresponds to a steady state value of effector T cells. This is particularly useful, as if the same pathogen strikes again, then the amount of antigen specific T cells is higher which would lead to a faster and more effective suppression of the pathogen.

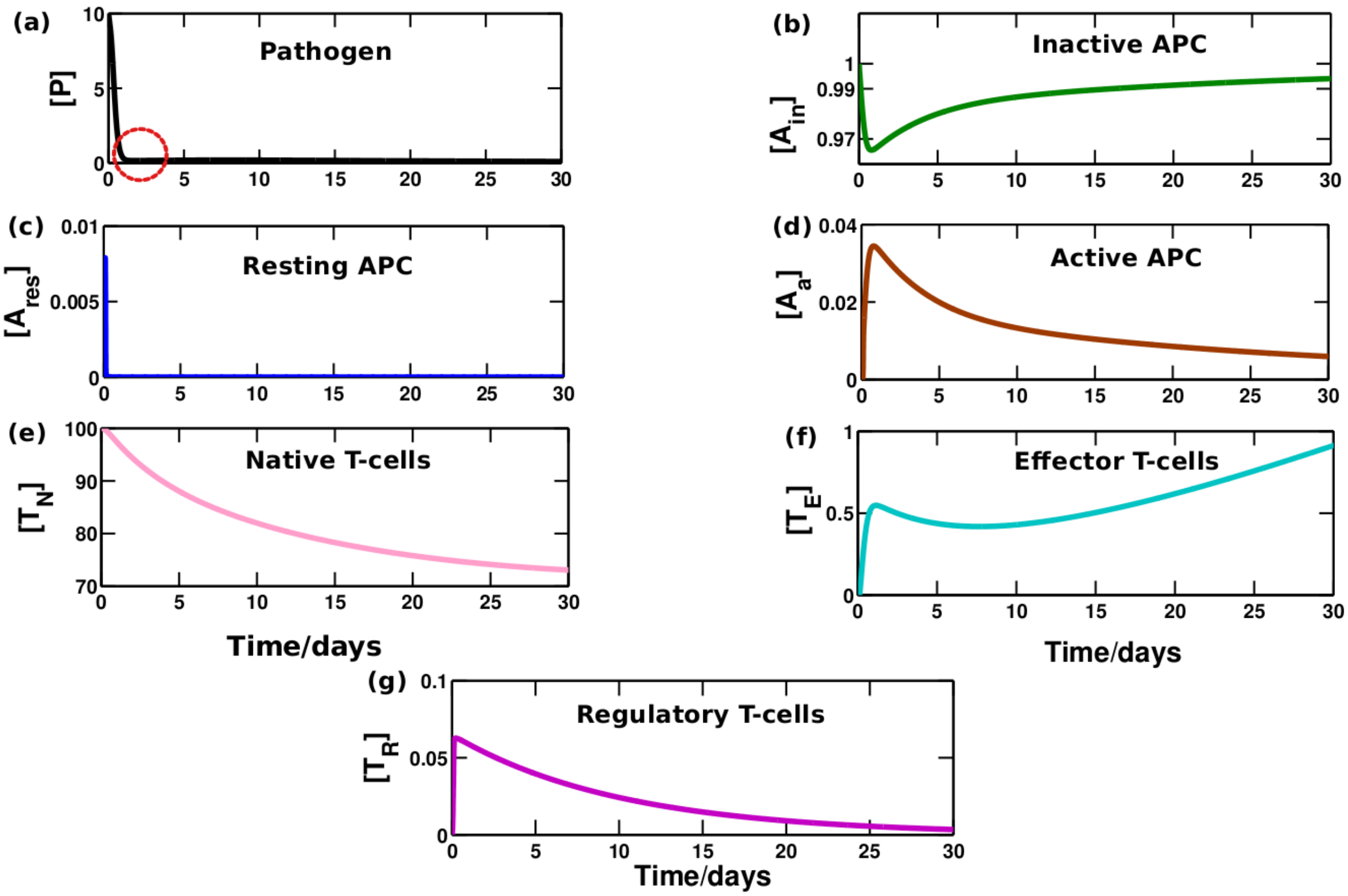

**Figure 4: Time evolution of immune response in absence of vitamin D.**

The dynamical variation is calculated in the weakly regulated state as mentioned in Figure 3 in absence of vitamin-D. Adaptive response sets in around 2 days after the pathogenic incursion. Pathogens are destroyed around the day 3-4 when effector T-cells reaches its peak value. Although regulatory T-cells control a little the huge constant production of T-effector cells, its steady concentration still preserves a significantly large value (0.5) which may increase the risk of autoimmune diseases. (Basic value parameters: $k_{inP} = 0.01, k_{aE} = 10^4$, other values are taken from Table 1)

**In the presence of standard level of Vitamin-D**

Vitamin D plays a crucial role in the onset of adaptive response. Nevertheless it modifies the scenario as we have explained in the last subsection. Once the T-effector population starts increasing, production of [D*] is upregulated (see **Figure 5**). This in turn upregulates Treg growth (initiated at around day 3), which along with [D*] regulates the aggressive, inflammatory responses exerted by T-effector cells, restoring order to the body. As the T-effector population decreases (after day 5), the [D*] concentration also starts dipping after reaching a maxima. However steady state has been reached at around100 days (see **Figure S2** in the supplementary material).

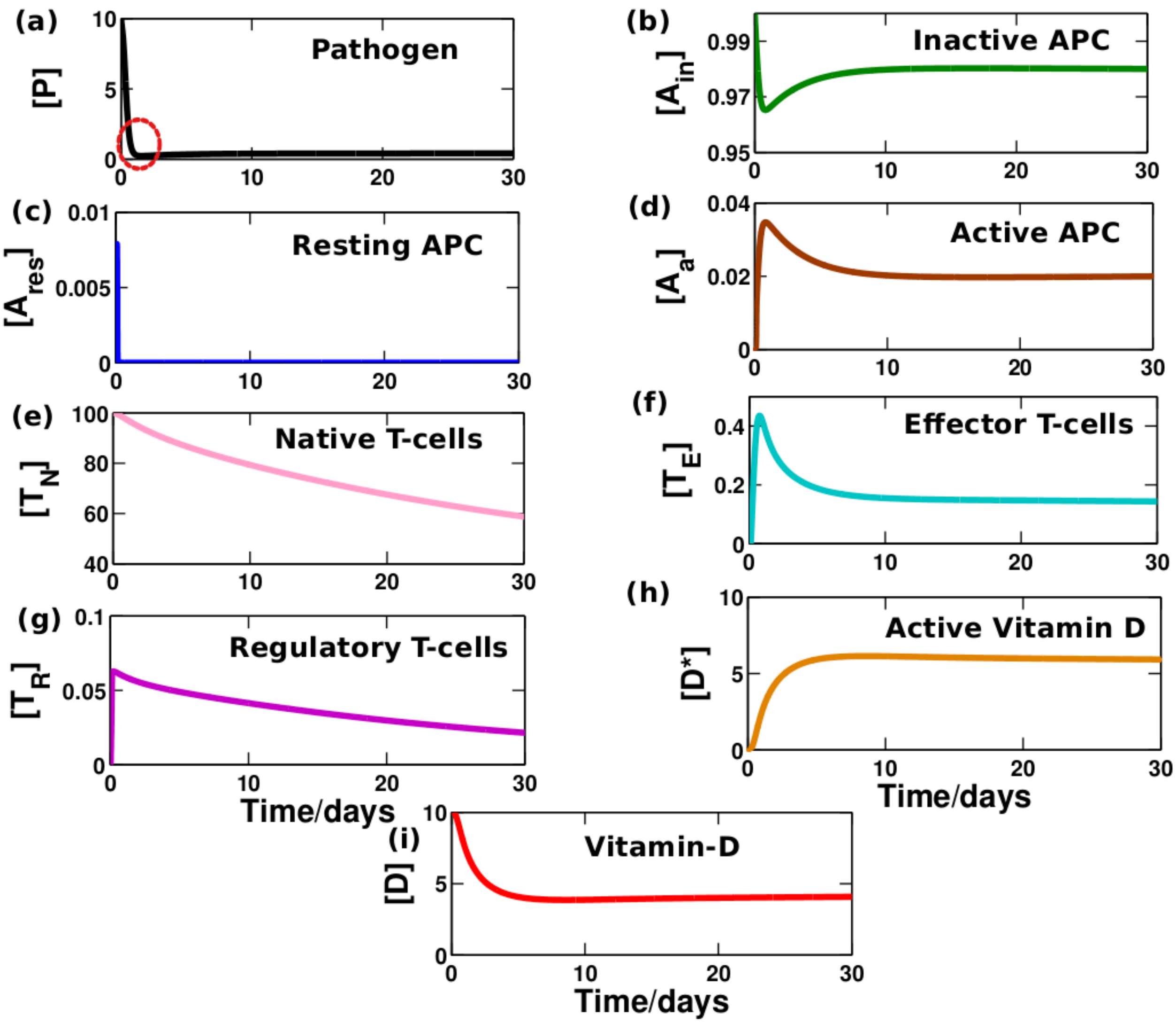

**Figure 5: Time evolution of immune response in presence of vitamin D.**

The dynamical variation is calculated in the weakly regulated situation as mentioned in Figure 3 but here in presence of vitamin-D. Adaptive response sets in around 2 days after the pathogenic incursion and here also pathogens are destroyed around the day 3-4 when effector T-cells reaches its peak value. Subsequently active vitamin D attains its peak value to control over the T-effector cells. Along with the enhanced regulatory T-cells production, active vitamin-D itself now control over the large production of T-effector cells. The steady state concentration of T-effector cells now low down significantly to a value of 0.2 which may decrease the risk of autoimmune diseases.

## Vitamin D and tolerance

A healthy immune system is always characterized by tolerating certain extent of pathogenic stimulation. The fact that vitamin D has been implicated as an important factor in several different autoimmune diseases suggests that vitamin D might be an essential element that normally participates in the control of self-tolerance. [39] From the model we explain that how the presence of vitamin-D can assist to attain a strongly regulated state that can reduce the risk of autoimmunity. We have considered here the moderately regulated/bistable regime as described in **Figure 3(e)** and **3(f)**. It is worth to mention here that experimental observations related to the adoptive transfer of tolerance also supports the emergence of such bistability where the balanced co-existence of strong and weakly regulated immune response preserve in the system. [40] Note that while strongly regulated state critically referred to system's tolerance for a certain pathogenic stimulation, weak regulation always mount system's response against that pathogen.

Weak or strong regulation crucially depends on the growth rate of T-effector as well as regulatory T-cells population. However the steady state population of Treg or T-effector cells is inversely related to their death rates. To investigate how the death rate of effector T-cells $(m_E)$ and death rate of regulatory T-cells $(m_R)$ mutually determine the mode of regulation we have varied both of them simultaneously. We have performed similar analyses as **Figure 3** to find out how a regulation is tuned between strong and weakly regulated states. Here in **Figure 6(a)** we observe that in absence of vitamin D, system is more tolerant towards the low death rate of regulatory T-cells which leads to a more steady state population of regulatory T-cells. However in presence of vitamin D as shown in **Figure 6(b)** we find a considerable enhancement of the tolerance level. Vitamin D still maintains the tolerant response even though there is high death rate of regulatory T-cells.

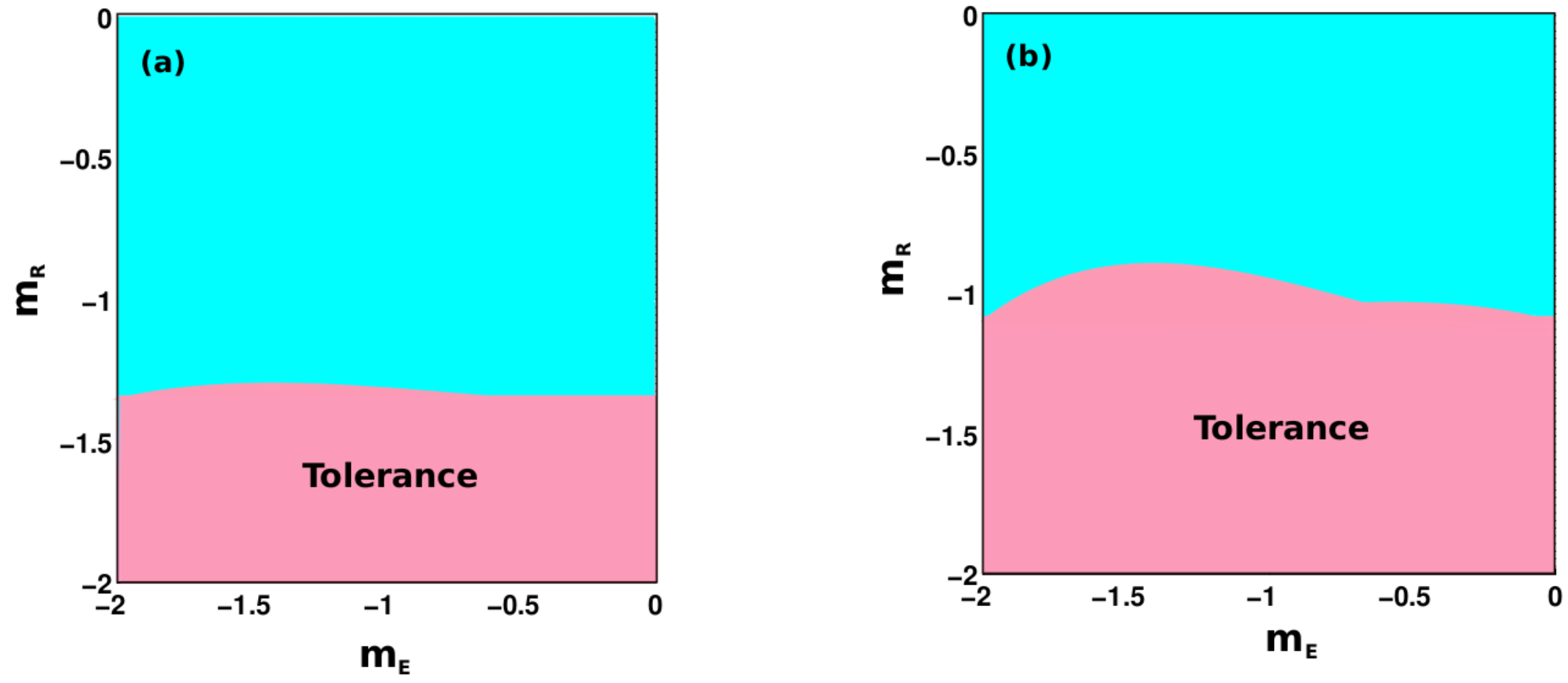

**Figure 6: Vitamin D induced tolerance level (emergence of strong regulation).**

Here we have varied both death rate of effector T-cells $(m_E)$ and death rate of regulatory T-cells $(m_R)$ to study the regulation in terms steady state concentration of regulatory and effector T-cells. Basic value parameters are taken as in given Figure 1(c) (also see Table 1) so that we are in the bi-stable regulation region. Initially, there are no effector T cells. Here pink zone correspond to the tolerance, i.e. the system falls into the strongly regulated state. Cyan zone corresponds to the development of a strong immune response, i.e. the system falls into the weakly regulated state. (Basic value parameters: $k_{inP} = 0.01, k_{aE} = 10^3$ , other values are taken from Table 1).

## Steady state analysis and optimal vitamin D

Similar to immune network theory [41] we have also assumed that the system has a large number of stable steady states that depends on several conditions that includes pathogen concentration, strength of stimulation, vitamin D concentration, APC activation rate and other various transition probabilities. Nevertheless the system can switch between two different steady states as has been often observed experimentally. For instance, low or high doses of an antigen can cause the system to switch to a suppressed state for the antigen, while intermediate doses can cause the induction of immunity. However in any particular condition system ultimately reaches to its steady state.

In steady state, for the present case, the body is assumed to be under the following constraints:

No Pathogenic incursion, so the entire pathogen term drops to zero. There are no T-effector cells. There are no $A_{res}$ or $A_a$ , i.e. mature APC cells. All DC's are immature or inactive.

Under the above constraints, the system of equations reduces to, at steady state:

$$\frac{dA_{in}}{dt} = \sigma_{in} - k_{inP} A_{in} T_E - m_a A_{in} = 0$$

$$\frac{dT_N}{dt} = \sigma_N - k_{aN} A_a T_N - k_{resN} A_{res} T_N - m_N T_N = 0$$

$$\frac{dD}{dt} = \sigma_D - k_{ED} T_E D - m_D D = 0$$

Solving these equations simultaneously, we get the following relations:

$$A_{in,ss} = \frac{\sigma_{in}}{m_a}$$

$$T_{N,SS} = \frac{\sigma_N}{m_N}$$

$$D_{ss} = \frac{\sigma_D}{m_D}$$

From the steady state relations, we can draw some important observations that make relevance to the state of the immune system. As assumed before, since all rate constants are time-invariant, we can see that the only variables are the production rates i.e. $\sigma$ values and death rates i.e. $m$ values that can only modify the steady state concentration of the precursor elements. However in absence of pathogen in the present case, we obtain the steady state concentration value for vitamin-D is 10.

## Vitamin D as a modulator

One important detail which needs to be addressed is the apparently new steady state in presence of pathogen in production of vitamin D with its tightly controlled homeostasis. To understand the relevance of vitamin D in the above response, different initial concentrations of vitamin D were taken to investigate its role. We have basically considered the initial concentrations of vitamin D near the steady state value of vitamin-D.

A range of concentrations from $10^{-2}$ to $10^{3}$ were taken (initially normalized to 10 in analysis in previous section) and the pathogen levels, T-cell levels and APC levels were noted in the newly established steady state. Steady state concentration of each element is plotted versus log([D ]. Very interesting results are observed (see **Figure 7**).

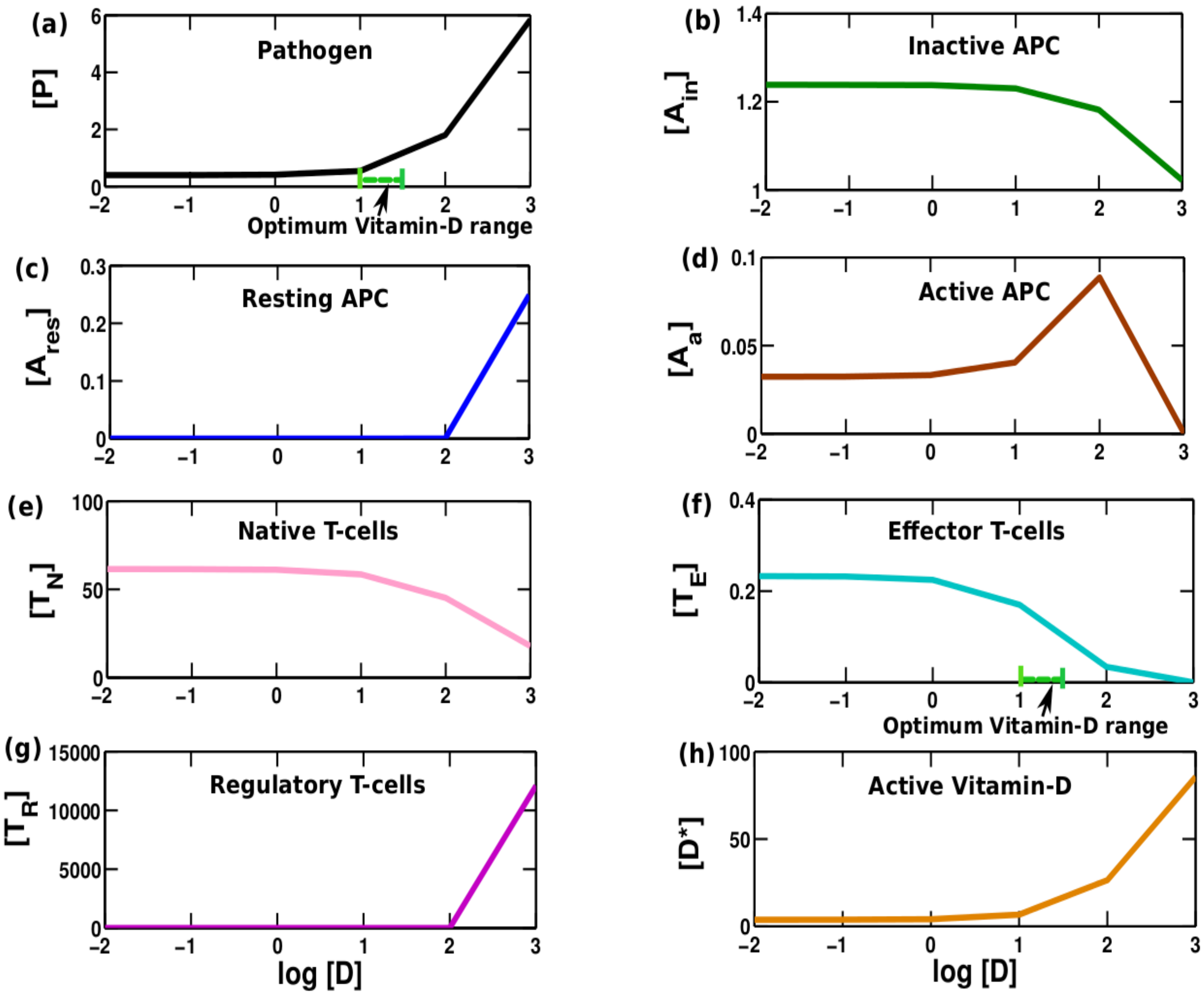

**Figure 7: Steady state value analyses as a function of log(initial vitamin D levels).**

At certain range of vitamin D concentration (around the value vitamin-D concentration 10) system gear up a more strong regulation where the steady state value of T-effector cells start on deceasing even with a faster rate. As a result pathogen concentration makes a sudden move towards its higher value to reappear back to the system. Note that at around vitamin-D concentration 100, each component shows a blast i.e., the concentration of the components decreases or increases very fast. However we indicate (with a green limit bar) the optimal vitamin-D ranges around 10-50 where both pathogen and effector T-cell level remain at reasonably low value. Vitamin-D level 100 corresponds to an alarming concentration compared to the standard vitamin D limit.

We indicate the optimal Vitamin-D level around 10-50 where both pathogen and effector T-cell level remain at reasonably low value. Recently a large number of epidemiological studies and an U.S. Institute of medicine committee start to shed light on various studies related to vitamin D. [42, 43] They reported that a serum 25-hydroxyvitamin D level of 20 ng/mL (50 nmol/L) is desirable for bone and overall health. Those studies indicated both the upper and lower limit of vitamin D intake. The high IgE levels were seen at very low 25-hydroxyvitamin D3 (<25 nmol/L) and at very high 25-hydroxyvitamin D3 (>135 nmol/L) levels. [44]

If we quantify the optimal level of vitamin D in the unit of nmol/lit, we are in the right ball park number.

We study the variation of each element around this optimal range.

**Pathogen:** We see that there is no/negligible amount of pathogen remaining for all concentrations of vitamin D below standard levels. This is because, less the vitamin D, less regulatory the system is and hence more aggressive is the immune response. This leads to complete eradication of pathogen. As vitamin D levels start being higher, the regulatory mechanism is highly upregulated leading to a more tolerant response. This can be noticed from the fact that beyond homeostatic standard levels, pathogen concentrations are not tends to zero and grow as a function of log [D].

**APC:** Below homeostatic levels, we see that the concentrations of inactive, active and resting APCs are largely invariant and constant. As we increase the concentration of vitamin D near the optimal level (100) the active APC reaches to its peak value. This is because there is no inhibition of activation of APC from very low concentrations of vitamin D. This suggests that there is constant T-cell activation, leading to a more inflammatory response at lower concentrations of vitamin D. At higher concentration of vitamin D, there is strong inhibition of active APCs and a strong decline of active APC concentration is seen. In contrary beyond the optimal level of vitamin D regulatory T-cell grows rapidly as function of log [D]. This offers a more tolerant response as the ability to activate T-effector cells is lesser. The result again suggests that at low vitamin D concentrations, inflammatory responses are more likely, which is a plausible cause for an increased risk of autoimmune response.

**Naive T cell:** The switch like behavior around homeostatic concentrations is noted. The reason for higher antigen specific T cells at lower vitamin D concentrations would indicate a higher probability of inflammation at low vitamin D levels. This is again in tune with the experiments that suggest vitamin D deficiency may be linked to autoimmune responses.

**T-effector cell:** Since T-effector cells are obtained from the activation of a naive T cell by an active APC, we see that the populations behave similar to those of native T-cells. The reasoning would be the same as for the APCs.

**T-reg cells:** Near the optimal range of vitamin D, an elevated level of regulatory T cells is buttoned on. This suggests that at the optimum vitamin D level there is an increased localized production of [D*] which serves to upregulate population of regulatory T cells by the time adaptive response sets in.

## Sensitivity to the values of parameters

### Scanning of Vitamin-D related basic value parameter

Because of the lack of information of the precise values of vitamin-D related rate constants it is crucial to investigate both the robustness and the sensitivity to these values. An additional reason for this study of the sensitivity is that these values change from system to system (here person to person) and the values can fluctuate even for the same person depending on the various conditions.

As both the T-cells i.e. Treg and T effector cells are modulated by the impact of active vitamin-D we have investigated the outcome of different possibilities of the combination of $k_{RD*}$ and $k_{ED*}$. Here we have scanned their parameter space to find the safe zone where both the pathogen and effector T-cells concentration remains under control. However at high vitamin D impact parameter pathogen may largely enhance their growth. Here $(\log k_{ED*}, \log k_{RD*})$ parameter space suggests that we should not exceed the value of $\log k_{RD*} = 0.1, \log k_{ED*} = 0.1$ to attain a safe zone (see **Figure 8(a)**). However $(\log k_{ED*}, \log k_{ED})$ parameter space suggests that we should stay behind the region where $\log k_{ED} = 1, \log k_{ED*} = 2$ to avoid high pathogenic raise up (see **Figure 8(b)**).

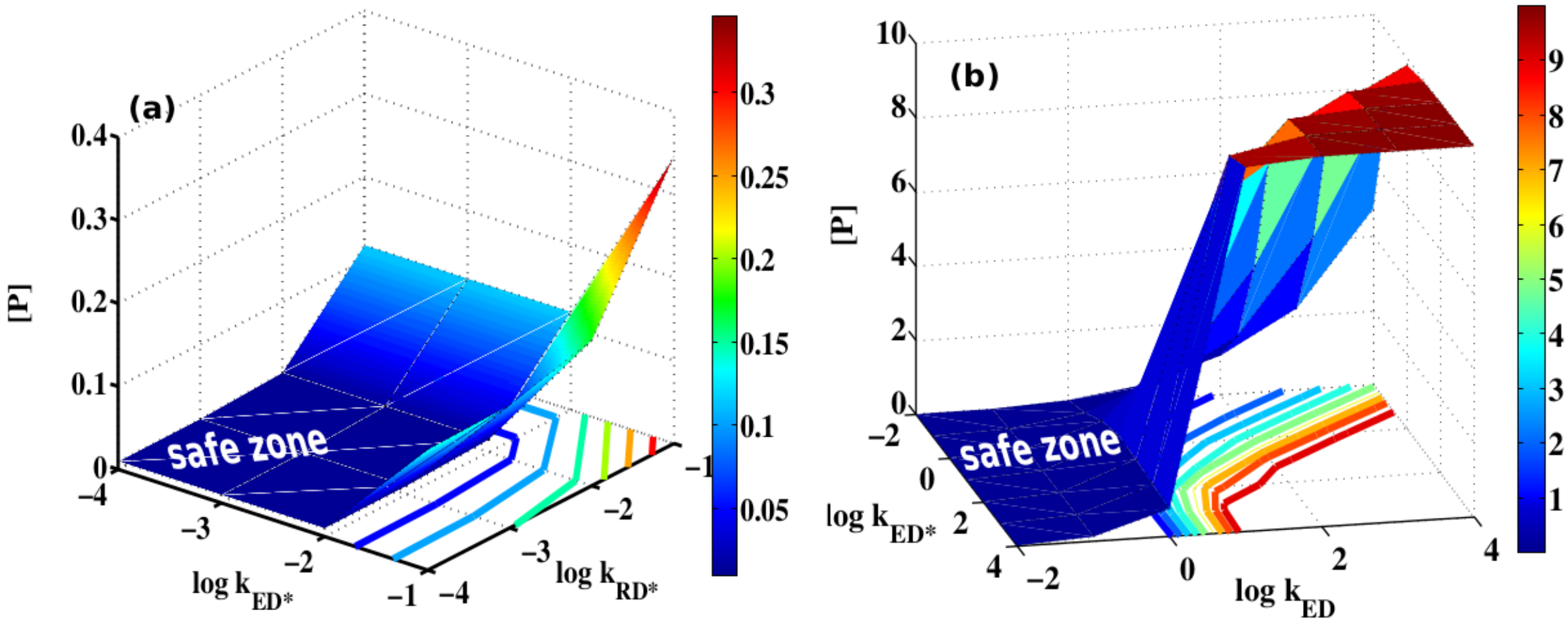

**Figure 8: Basic value parameter scan.**

Investigation of vitamin-D impact- (a) Over regulatory T-cells up-regulation $(k_{RD^*})$ and effector T-cells down-regulation rate $(k_{ED^*})$, simultaneously in their parameter space, and (b) over effector T-cells down-regulation rate $(k_{ED^*})$ and effector T-cell initiated active vitamin-D production rate $(k_{ED})$ simultaneously in their parameter space. $(\log k_{ED^*}, \log k_{RD^*})$ parameter space suggests the value $\log k_{RD^*} = 0.1, \log k_{ED^*} = 0.1$ is the alarming zone beyond which pathogen may retain in the system. $(\log k_{ED^*}, \log k_{ED})$ parameter space suggests the value $\log k_{ED} = 1, \log k_{ED^*} = 2$ is the alarming zone beyond which effect of vitamin-D is carcinogenic that allow reappearance of a large amount of pathogen. Here we have taken weak pathogen stimulation $(k_{aP} = 0.0001)$ while other parameter values are taken from Table 1.

## Weak simulation and strong regulation: A specific regime

Each day each moment immune system protects us against millions of hazardous foreign elements. Whether the pathogenic stimulation strong or weak a healthy immune network always maintains a balanced regulation over the T-effector cells concentration to reduce the risk of autoimmunity. We are aware of the fact that while presence of vitamin-D at its standard level is required to control the effector T-cell explosion, above a threshold vitamin-D concentration causes reappearance of the pathogen above the tolerance level of our immune system. To investigate such possibilities we have evaluated the pathogen concentration varying the negative impact of [D*] over APCs $(k_{aD*})$ and effector T-cells $(k_{ED*})$ simultaneously. We find interestingly three regions (**see Figure 9(a)**) (i) safe zone at low vitamin-D impact, where pathogen concentration is found to be negligible, (ii) pathogen reappearance zone, where initially pathogen concentration becomes lower by the peak value concentration of effector T-cells. However at longer time pathogen and effector T-cells shows a damping oscillation but both waves move in an anti-correlated fashion and finally reach their steady state values. (see **Figure 9(b)**). This oscillation supports the normal tolerance to pathogen and the remaining T-cells preserves the long term memory of the past attack. These combinations support the regular defense mechanism by which our immune system works. (iii) In the third region we find a pathogen sustaining zone where at high vitamin-D impact system is hyper regulated and steady state concentration value of pathogen remains significantly large.

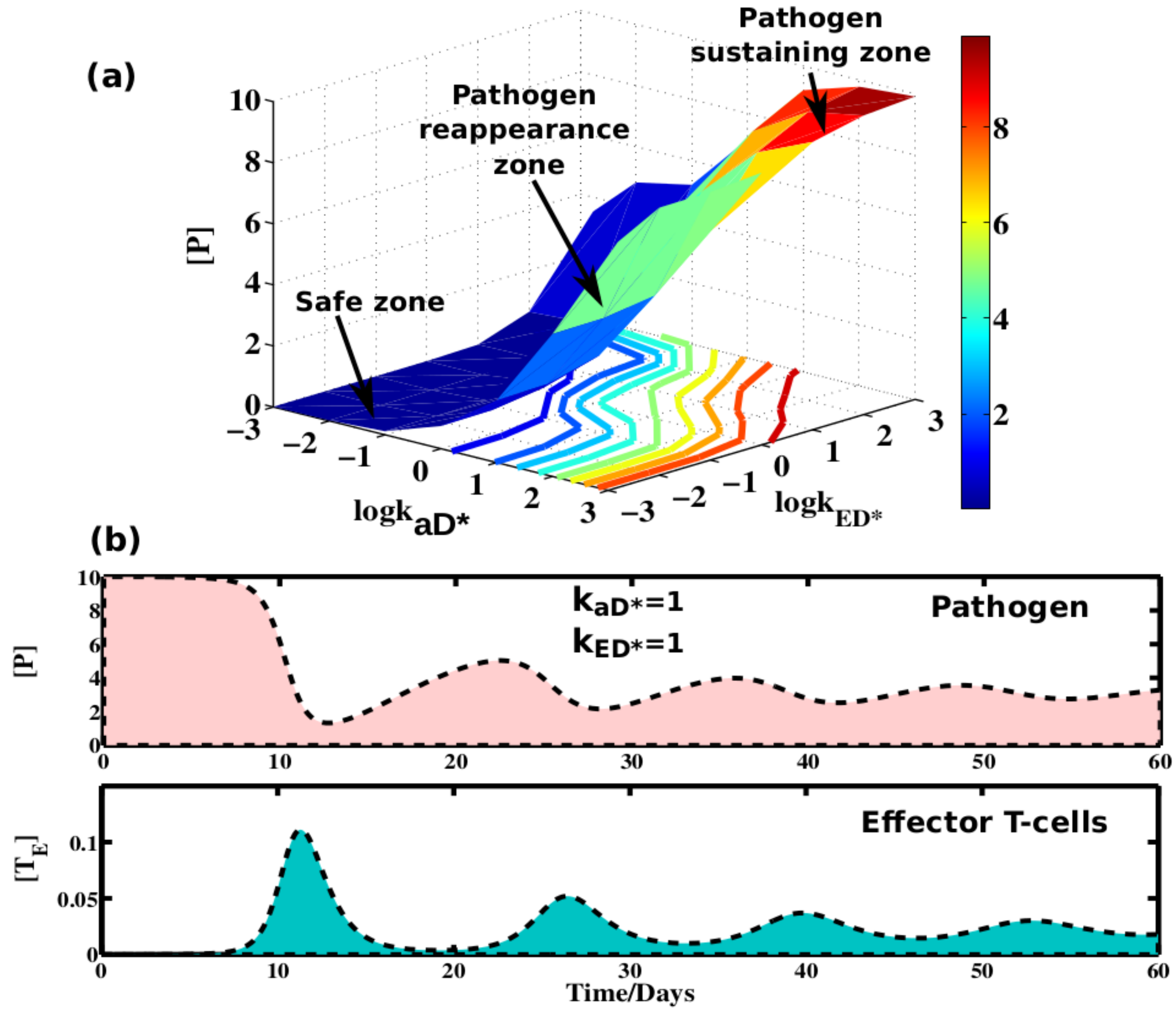

**Figure 9: Impact of vitamin-D over pathogenic profile.**

(a) We vary simultaneously the impact of [D*] over APCs $\left(k_{aD*}\right)$ and effector T-cells$\left(k_{ED*}\right)$ . We find three different regions: safe zone where pathogens are low in concentration. Reapperance of pathogen occurs at moderate vitamin-D impact imposed on active APC and effector T-cells. At high vitamin-D impact steady state concentration of pathogen largely increases. (b) At pathogen reappearance zone effector T-cells and pathogen propagates with a damping oscillation and reaches to their steady state values at longer time. Here we have taken weak pathogen stimulation and strong effector T-cell regulation $\left( k_{inP} = 0.0001, k_{aE} = 10 \right)$ while other parameter values are taken from Table 1.

## Discussion

Vitamin D, the "Sunshine Vitamin" is now well recognized for the immunomodulatory functions beyond its classical function in calcium metabolism. To understand the immune regulation exerted by vitamin D in this study we develop a theoretical coarse-grained model based on interaction network. The network dynamically connects different immune components that are involved in the vitamin D regulated immune responses. The formulated kinetic scheme describes time evolution of these components that include pathogen, APCs, effector T cells, regulatory T cells, etc. The steady state analyses of the proposed master equations suggest intricate relations between vitamin D levels and regulatory T cells maintained by homeostasis where vitamin D play role as a strong regulatory activator. We found quantitatively that lower levels of concentration of active vitamin D ([D*]) (lower than optimum) correspond to weak regulation. In this limit once a pathogen/antigen enters the body, the nature of the immune response would be more inflammatory. This eventually leads to autoimmune diseases. On the other hand our model predicts that vitamin D enhances the tolerance level of immune system, thereby increasing resistance to auto-immune diseases. Quantitative predictions of the model are in good agreement with several other experimental studies and clinical observations. Our model essentially attempts to quantify how much vitamin D is needed to resist autoimmunity and why.

The steady state analysis of the proposed master equations revealed intricate relations between vitamin D levels and T-regulatory cells maintained by homeostasis. These relations suggested that at homeostasis, lower levels of [D*] correspond to a lower population of T-regulatory cells, which suggests that once a pathogen/antigen enters the body, the nature of the immune response would be less regulatory and hence more inflammatory(aggressive). The simulations seem to be in fair agreement with the typical progression of an immune response.

The critical role of the various cells, especially localized [D*] concentration is understood via investigating the dynamics of the response. This was done through simulating the coupled differential equations. Further, there exists a delicate window of concentrations of vitamin D which would be critical in maintaining an appropriate response to a pathogen. Extremely low levels of [D*] could lead to increased risk of autoimmune responses and extremely high levels would suggest an extremely tolerant response, which could increase the risk of tumors and cancerous cell growth.

It is important to note that two enzymes CYP27B1 and CYP24A1 and the population of VDR play important role in balancing several immunological responses. Our coarse-graining model would predict that defect/unavailability of any of these proteins will greatly perturb the whole system. A series of D*-VDR mediated processes that have enormous consequences have not been fully understood yet. Malfunction of these enzymes (such as: CYP27B1 and CYP24A1) can also reflect a deeper problem that is difficult to rectify.

It is worth mentioning here that the activation of a naive T cell into an effector or regulatory T cell is a complex process. This begins with the scanning of the surface of the APC's in the lymph nodes for the MHC class II type molecules by the naive T cells. If a particular epitope is recognized and co-stimulatory molecules are present, then the activation process is initiated. Now this can be understood via an energy landscape analysis. The process of successful activation can be thought of as the T cell negotiating a barrier in the energy landscape. This can be brought about through either a single successful contact with an APC or multiple contacts if the second or later contact occurs within a finite time. If the T-cell is above the seperatrix in the energy landscape then the probability of a successful activation is higher which is only present for a finite time after the previous excitation. This is very similar to the immunological studies carried out by Hong et al. [45] and Chakraborty et al. [46] and the enzyme catalysis model proposed by Min, Xie and Bagchi earlier. [47] However to make the present model tractable, we had to ignore such complexity of T-cell activation.

The master equation approach adopted here gives a deterministic description of the problem, given the initial values of the parameters and the fluxes. Within a biological cell, there can always be large fluctuations due to environmental factors or other causes. [48, 49] Such fluctuations can induce cross-over from weak regulation to strong regulation.

In future, we plan to extend our system of equations to include effects of drugs such as immune suppressants such as glucocorticoids that introduce a further competition in the reaction network of the body. The role of immune suppressants is not understood but known to exhibit serious side effects like cancer.

## Acknowledgments

It is a pleasure to thank Prof. Anjali A. Karande, Ms. Ritu Mishra, Ms. Gauri Ranadive, Mr. Manoj Kumar Mahala and Mr. Kushal Bagchi for helpful discussions. This work was supported in part by grants from DST and BRNS, India. BB acknowledges support from a Sir J.C. Bose Fellowship, DST.